\documentclass[acmsmall,screen]{acmart}

\setcopyright{acmlicensed}
\acmJournal{PACMHCI}
\acmYear{2023}
\acmVolume{7}
\acmNumber{CSCW2}
\acmArticle{292}
\acmMonth{10}
\acmPrice{15.00}
\acmDOI{10.1145/3610083}

\usepackage{listings}

\begin{document}

\title{Rethinking People Analytics With Inverse Transparency by Design}

\newcommand{\vzaffil}{%
	\institution{Technical University of Munich}
	\city{Munich}
	\country{Germany}%
}

\author{Valentin Zieglmeier}
\affiliation{\vzaffil{}}
\email{valentin.zieglmeier@tum.de}
\orcid{0000-0002-3770-0321}

\author{Alexander Pretschner}
\affiliation{\vzaffil{}}
\email{alexander.pretschner@tum.de}
\orcid{0000-0002-5573-1201}

\authorsaddresses{%
	Authors' addresses: Valentin Zieglmeier, valentin.zieglmeier@tum.de; Alexander Pretschner, alexander.pretschner@tum.de, Technical University of Munich, TUM School of Computation, Information and Technology, Chair of Software and Systems Engineering, Boltzmannstr. 3, 85748 Garching, Germany%
}


\begin{abstract}
	Employees work in increasingly digital environments that enable advanced analytics. Yet, they lack oversight over the systems that process their data.
	That means that potential analysis errors or hidden biases are hard to uncover.
	Recent data protection legislation tries to tackle these issues, but it is inadequate.
	It does not prevent data misusage while at the same time stifling sensible use cases for data.

	We think the conflict between data protection and increasingly data-driven systems should be solved differently. When access to an employees' data is given, all usages should be made transparent to them, according to the concept of \emph{inverse transparency}.
	This allows individuals to benefit from sensible data usage while addressing the potentially harmful consequences of data misusage.
	To accomplish this, we propose a new design approach for workforce analytics software we refer to as \emph{inverse transparency by design}.

	To understand the developer and user perspectives on the proposal, we conduct two exploratory studies with students.
	First, we let small teams of developers implement analytics tools with inverse transparency by design to uncover how they judge the approach and how it materializes in their developed tools.
	We find that architectural changes are made without inhibiting core functionality.
	The developers consider our approach valuable and technically feasible.
	Second, we conduct a user study over three months to let participants experience the provided inverse transparency and reflect on their experience.
	The study models a software development workplace where most work processes are already digital.
	Participants perceive the transparency as beneficial and feel empowered by it.
	They unanimously agree that it would be an improvement for the workplace.
	We conclude that inverse transparency by design is a promising approach to realize accepted and responsible people analytics.
\end{abstract}



\maketitle

\newcommand{\citeauthorcite}[1]{\citeauthor{#1}~\cite{#1}}

\hyphenation{tool-chain}


\section{Introduction}
\label{sec:introduction}

Workplaces are becoming increasingly digital.
Everything from employee communication to the status of tasks and work items is stored and handled digitally.
This enables advanced people analytics that process employee data to speed up processes or help with decision-making.
But, contrary to the consumer context, employees often have no choice which tools to use in their workplace.
Furthermore, the data processing is opaque to those subjected to it.
This lack of control and oversight by employees raises concerns~\cite{teebken2021privacy} and means that there is little recourse in case of data misusage or discriminatory analysis errors.
With increasingly automated and automatic decision-making, the risks of data misusage rise further~\cite{lustig2016algorithmic, giermindl2022dark}.
Data-based insights can play a role in deciding if a person should be invited for a job interview, if they should be assigned to a project, or qualify for a promotion~\cite{tursunbayeva2021ethics}.
It is therefore vital that any discrimination or misusage of data can be uncovered and challenged~\cite{rieke2018public}.

To protect individuals' privacy and ensure accountability, data protection legislation is employed~\cite{rubinstein2010privacy, flannery2017gdpr}.
Depending on the cultural context and underlying trust model, it takes different shape~\cite{bowie2006privacy, pretschner2014achieving}.
A traditional approach is \emph{detective enforcement}, which relies on self-regulation and voluntary codes.
It assumes many data usages to be benign and enables them by default.
Terms of use or non-disclosure agreements enable \emph{ex post} reaction to misusage~\cite{povey1999optimistic, pretschner2014achieving}.
This optimistic solution is common in many parts of the United States of America~\cite{rubinstein2010privacy}.
Recent privacy legislation such as the 2016 General Data Protection Regulation (GDPR)~\cite{eu2016gdpr} of the European Union and the 2018 California Consumer Privacy Act (CCPA)~\cite{cali2018privacy} goes beyond voluntary codes to implement formal privacy regulation and provide individuals more control over their data~\cite{bowie2006privacy, rubinstein2010privacy}.
They require the implementation of technical measures that prevent extraneous data usage.
This can be seen as a move towards \emph{preventive enforcement}, which only permits data usage for purposes specified in advance.
Thereby, it strives to prevent misusage \emph{ex ante}~\cite{vanbeek2007comparison, pretschner2008usage}.

We think this increased protection is important, but in many cases insufficient to prevent data misusage in the workplace, while at the same time stifling sensible use cases for data.
Four factors are, in our view, mainly responsible for this.
To start with, (1)~opting out of data sharing is not always possible for employees.
If the data processing is necessary for core business processes, it does not require consent.
And the power asymmetry between employees and management and often forced usage of digital technologies in the workplace mean that, even if employees legally have the right to object, denying consent may effectively remain a theoretical possibility.
In case there is a choice, though, (2)~use cases for data are becoming more complex, making it harder for individuals to fully understand the impact of giving access to their data.
The lengths of typical privacy policies\footnote{Google's for example, when viewed as a PDF, is 30 pages long: \url{https://www.gstatic.com/policies/privacy/pdf/20210701/7yn50xee/google_privacy_policy_en_eu.pdf} (last accessed 2022-01-20)} show the complexity of data processing, meaning a full understanding is questionable \cite[see also][]{mcdonald2008cost, tang2021defining}.
This is exacerbated by the fact that (3) software tools are not static products.
Software-as-a-service and agile programming mean that software evolves continuously~\cite[pp.~19--20]{gurses2018privacy}.
Even if individuals had the capacity to understand how data are processed by their employer, their knowledge can therefore quickly become obsolete.
Finally and importantly, (4)~a blanket decision for or against data sharing cannot always be made, as the usage context is an important decision factor.
Data that are given away can be used in \emph{unexpected} and \emph{unintended} ways~\cite{rao2016expecting, hummel2018sovereignty}, and hence be misused from the perspective of the data subject.
This can happen intentionally, by trickery or hiding of essential information, or unintentionally, by misreading or misrepresenting those data.
Faced with these concerns, overwhelmed by choice and a lack of oversight, and backed by laws such as the GDPR and CCPA, employees might therefore aim for absolute data minimization to lower perceived risks.
This ideal certainly reduces the potential for misusage, but opting out could lead to other disadvantages, such as lack of access to data-driven features. 
In addition, it becomes difficult for companies to utilize data beyond cases in which they are absolutely necessary, restraining legitimately helpful data usages and stifling research and innovation in the big data space~\cite{jia2018effects, zarsky2017incompatible, gal2020competitive}.

We think that this issue can be solved differently, drawing inspiration from \citeauthorcite{brin1998transparent}.
They describe a dystopia in which citizens are monitored by the police during their every move, making their lives transparent. To balance this, they propose to empower individuals by letting them watch over their watchers---providing what they call \emph{inverse transparency}.~\cite{brin1998transparent}
This idea has been previously proposed as a new digital leadership concept, aiming to solve tensions between managers and employees~\cite{gierlich2020more}.
Continuing these thoughts, we envision making all usages of employees' data visible (transparent) to them.
We think that getting an overview of how their data are used empowers individuals to gain true \emph{data sovereignty}~\cite{hummel2018sovereignty}, meaning ``self-de\-ter\-mi\-na\-tion [...] with regard to the use of their data''~\cite[p.~550]{jarke2019data}.
This has the potential to improve their trust to allow data usages that might be beneficial to them.
In addition to helping to uncover misusage retroactively which enables \emph{accountability}~\cite{weitzner2008information}, this transparency could also increase the \emph{felt accountability} of data consumers~\cite{hall2017accountability}, thereby deterring data misusage even before it occurs.

Some data, such as health or genetic data, will always warrant preventive enforcement due to their high sensitivity.
Furthermore, due to the power asymmetry in the workplace, the added transparency alone is not sufficient to protect individuals.
Additional safeguards, such as strong workers' councils or appropriate recourse in case of data misusage~\cite[see, e.g.,][p.~36]{mundie2014privacy}, may therefore be a prerequisite for this idea.
Given those, we think it could be a promising solution to the conflict between data protection and data-based use cases.

To concretize our vision, we consider the example of software developers in an IT company.
These employees can work remotely, with some companies even adopting ``all remote'' configurations~\cite{choudhury2020gitlab}.
This can increase the employer's interest in monitoring employees with people analytics.
As software developers are in high demand in the labor market, though, the inherent power asymmetry between them and their employer is reduced.
Data about these employees are stored in various systems and accessed through a multitude of tools.
In this scenario, employees track their work in issue tracking software and use a workplace messenger.
That means that data exist about the specific technologies and problems they work on, as well as whom they collaborate with.
The traditional detective enforcement allows utilizing these data for, e.g., managerial decision-making or collaboration between colleagues.
However, it makes room for profiling and patronization of employees based on data that might not represent the full picture or be inadequate for these uses.
Employees might be fired or discriminated against due to misinterpreted or misused data, and have no recourse against it.
With preventive enforcement on the other hand, any data usage beyond those required by core work processes is forbidden.
This makes it difficult to implement systems enabling advanced data-based use cases.
Yet, as we have deliberated above, misusage of data is not sufficiently prevented.
If we now imagine the same example with the envisioned transparency over data usages, those issues are addressed.
Employees are free to collaborate without any overhead, and data can be utilized for company-level decision-making.
Should data be misused and harmful consequences for an employee arise, they have access to an audit trail.
To defend themselves, they can make it available to their workers' council or a lawyer to support their case.


In this paper, we explore the idea of enshrining inverse transparency into people analytics from their conception. We aim to understand how this could change software design and, by extension, foster employees' trust in and acceptance of sensible data usage processes.
Our goal is to facilitate data-based use cases that can be beneficial for individuals, while better protecting them from misusage of their data.
Evolving the idea of \emph{privacy by design}~\cite{cavoukian2009privacy}, we describe the software development paradigm of \emph{inverse transparency by design}.
As an empirical contribution, we conduct two exploratory studies with students in a controlled environment to understand the developer and user perspectives.
In the first, we explore the implications of our approach for software design. To that end, we let small teams of student developers implement various analytics tools based on the principles of inverse transparency by design.
We then analyze and discuss the changes they make to their tools to meet transparency requirements, and how they judge the approach in their reflections.
In our second study, we consider the perspective of data subjects on our concept. Therefore, we conduct a controlled laboratory study with students that mirrors the real-world use case of a software development department.
Participants worked for three months in a workplace-like setting utilizing transparency-enabled people analytics.
We examine and deliberate the user experience and personal perspectives of participants.
Note that, for both studies, we worked with university students in a controlled environment.
This was an intentional choice to best study a fundamental change in the work and interactions of employees.
Thereby, we remove potential confounding factors~\cite[see, e.g.,][]{bovey2001resistance, nan2009impact} to support internal validity~\cite{fiske2005laboratory} and establish causality~\cite{katok2018designing}.
The artificial nature of the studies may limit external validity, though.
We discuss the impact and our reasoning in more detail in \autoref{sec:discussion}.

In all, we contribute a comprehensive conceptualization and preliminary evaluation of inverse transparency by design, a new approach for workplace software development.
When studying the developer perspective, we find that building with inverse transparency by design does not inhibit core functionality, with developers considering the concept practical.
In our study of the user perspective, participants experiencing inverse transparency in practice find it beneficial and feel empowered by it. Given the choice, they would unanimously opt for it in their workplace.
We conclude that moving towards incorporating inverse transparency by design is a promising direction for people analytics.

\section{Related Work}
\label{sec:related-work}

We begin by describing other ideas besides inverse transparency that aim to find a middle ground between necessary data protection and sensible uses for personal data.
Then, we give an overview of related works that specifically propose to provide transparency over data usages to ensure accountability.
Finally, we narrow down further to the technical realization.
We discuss other works that aim to change personal data processing such that individual data usages are tracked.

\subsection{Balancing Data Protection and Sensible Data Usage}

The conflict between data protection and sensible data usage has been considered in various works.
\citeauthor{cate2002principles} observed that ``many uses of personal information pose no risk of harm to individuals, while creating significant benefits for data subjects and society more broadly''~\cite[p.~37]{cate2002principles}.
As a solution, they propose \emph{use-based privacy}, which entails defining collective norms that specify acceptable uses~\cite{cate2002principles, birrell2018sgx}.
For example, these could explicitly exclude discriminatory uses of data where they are obvious.
Then, the norms are enforced automatically by systems that process data~\cite[see, e.g.,][]{birrell2018sgx}.
A similar strategy underlies \emph{distributed usage control}, which instead of collective norms aims to enforce user-defined usage policies~\cite{pretschner2009overview} with comparable technical challenges~\cite[see, e.g.,][]{wagner2018distributed}.

Independently of who defines the usage policies, though, it is in our view impossible to know all acceptable usage patterns for data in advance.
Therefore, we consider it important to enable more flexibility and instead provide individuals with inverse transparency, allowing them direct oversight over how their data are used.
Potential misusage of data can then be handled retroactively, ensuring accountability.
Still, the idea to establish collective norms defining appropriate usage of data, enshrined for example in laws or company agreements, is compelling.
They could serve as a minimum safeguard for individuals, with inverse transparency helping to protect them for data usage that goes beyond the basic use cases.

\subsection{Providing Transparency Over Data Usages to Ensure Accountability}

Our work is not the first to introduce the idea of \emph{inverse transparency}, or more broadly aiming to ensure accountability by giving individuals oversight over usages of their data.
The general concept of inverse transparency was originally conceived of and presented by \citeauthorcite{brin1998transparent} (see \autoref{sec:introduction}).
In the software development context specifically, similar concepts have been developed, also aiming to provide transparency to ensure accountability.
An important predecessor to our work is the paper on \emph{hippocratic databases} by \citeauthorcite{agrawal2002hippocratic}. They discuss the usefulness of giving individuals access to audit trails of databases holding their information, allowing them to detect misusage~\cite{agrawal2002hippocratic}.
\citeauthor{weitzner2008information} also deliberate the potential benefits of making data usages transparent to individuals, achieving what they refer to as \emph{information accountability}~\cite{weitzner2008information}.
They see two main advantages over the status quo:
First, reducing individuals' mental load as they do not have to judge \emph{ex ante} all potential usages of their data.
Second, enabling redress in case of harmful misusage of data~\cite{weitzner2008information}.

These works serve as a motivation and theoretical foundation to our work.
We build on their ideas to propose a new approach for people analytics development: inverse transparency by design.
As our added contribution, we study its potential effects on software development and users.

\subsection{Changing Personal Data Processing to Integrate Usage Logging}

Finally, we discuss technical solutions related to our concrete idea how to ensure that employees are provided transparency over all usages of their data: by rethinking people analytics with inverse transparency by design.
People analytics today are typically designed the same way as any other business analytics: the data they operate on are collected elsewhere and presumed as given~\cite{huellmann2021it}.
In a sense, the tools consider these data as a mere resource, with employees under analysis reduced to data sources.
Data flow in one direction only: from employees to the tools that analyze them~\cite[see, e.g.,][]{mathur2015quantified, sap2020optimize, fabbri2022work}.
This is exemplified by how SAP visualizes the data processing pipeline for their SuccessFactors\footnote{\url{https://www.sap.com/products/hcm/workforce-planning-hr-analytics.html}} product, a large people analytics suite:
On a high level, they depict a one-way pipeline from various data sources, such as ``human resources,'' into their analytics platform~\cite[p.~3]{sap2020optimize}.
Considering that people analytics analyze humans and not business processes, though, this lacking consideration of individuals' interests has been critically reflected upon~\cite[see, e.g.,][p.~417]{giermindl2022dark}.

We propose to rethink people analytics design to, conceptually, add a reverse data flow into this process.
By integrating data usage tracking into the tools and sending the logs back to the employees under analysis, we give them a view into how their data are analyzed.
To our knowledge, we are the first to propose such a rethink of people analytics design.
Conceptually comparable ideas have been proposed previously, though.
As notable examples, \citeauthorcite{sundareswaran2012ensuring} and \citeauthorcite{bagdasaryan2019ancile} describe implementations of usage logging that could enable inverse transparency.
The \emph{CIA framework}~\cite{sundareswaran2012ensuring} is based on packaging personal data together with a usage logging module in Java JAR files before giving access to them~\cite{sundareswaran2012ensuring}.
\emph{Ancile}~\cite{bagdasaryan2019ancile} on the other hand is an online privacy platform. \citeauthor{bagdasaryan2019ancile} identify many of the same challenges that we see.
Their \emph{Ancile} server considers Python analysis scripts that are sent to it for execution in a sandbox~\cite{bagdasaryan2019ancile}.
Both implementations are relevant to learn from and can be seen as potential instantiations of inverse transparency in software.
Yet, they both consider neither the developer nor the user perspective.
On the one hand, understanding the developer perspective is important to judge the consequence of transparency requirements on software development.
Is it realistic to expect software to be written based on a new paradigm?
Could this requirement inhibit development of innovative features?
Previous research on privacy by design has recognized the importance of considering the developer perspective to judge the viability of software design principles, as their success directly depends on developers' actions~\cite[see, e.g.,][]{hadar2018privacy, senarath2019will}.
On the other hand, considering the user perspective is a vital aspect of privacy and empowerment.
While in theory the idea of making visible usages of data can sound obvious and like a clear benefit, we need to deliberate the effect of this transparency on individuals.
Does it foster their trust and increase their acceptance of people analytics?
Or could it lead to new concerns or other negative consequences for them?
To close these gaps, we present results from two empirical studies, which are our novel contribution compared to these works.

\section{Concept}

Currently, data usage processes in the workplace happen without oversight of the employees concerned by them.
To tackle this, we think that people analytics should be built with inverse transparency by design, a next step after privacy by design.
This means they should be built in such a way that data usages can be traced back and attributed.
As a first step, we discuss our idea as a theoretical concept, outlining the basic requirements for inverse transparency.
Second, we deliberate the potential implications for software design.
Finally, we discuss the problems of usage log integrity and system trustworthiness as necessary prerequisites for our concept to function.

From here on, instead of talking specifically about employees or managers, we will refer to the more generic concepts of \emph{data owners} and \emph{data consumers}.
This reflects that these roles might be reversed or even inhabited by the same person (accessing their own data).
We follow the definition by \citeauthor{pretschner2006distributed}, who state that for each datum, there is a \emph{data owner}~\cite{pretschner2006distributed}. They ``[possess] the rights to the data''~\cite[p.~40]{pretschner2006distributed}.\footnote{For example, in case the GDPR applies, this would correspond to the ``data subject,'' meaning the individual identifiable through the data~\cite[Art.~4]{eu2016gdpr}.}
They also define the role of the \emph{data consumer}~\cite{pretschner2006distributed}.
We personify the data consumer in our concept. In cases of algorithmic data usages, the processing system could be considered the data consumer.

\subsection{Requirements for Inverse Transparency}
\label{sec:concept}

The basis of inverse transparency is to give data owners an overview of all data usages. That requires (1) monitoring every usage of data, (2) verifying the authenticity of these events and storing them, and (3) making this information transparent to data owners.
According to the separation of concerns, we can imagine each requirement being fulfilled by individual tools, but the functionality can also be integrated into the operating system or directly into the software that provides data access.

To enable the three steps of our vision, let us therefore consider three (conceptual) components:

\begin{enumerate}
	\item \emph{Monitor}: Track data usages
	\item \emph{Safekeeper}: Store monitored usages
	\item \emph{Display}: Make stored usages transparent
\end{enumerate}

\begin{figure}[htbp]
	\centering
	\includegraphics[width=0.65\linewidth]{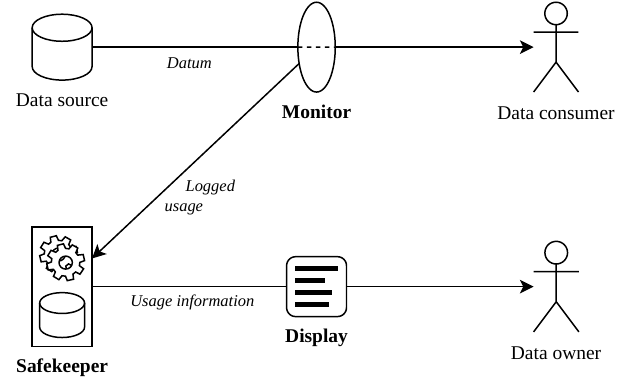}
	\caption{Providing inverse transparency on a conceptual level. The data consumer accesses a datum from the data source. Their usage is logged and stored. The data owner can now retrieve the logged usage information.}
	\Description{The conceptual components of the transparency framework are shown. From the Data source, an arrow points to the data consumer, labeled ``Datum''. Interposed is the unit ``Monitor''. From this, an arrow labeled ``Logged usage'' points to the unit ``Safekeeper''. From this, an arrow labeled ``Usage information'' points to the Data owner. Interposed is the unit ``Display''.}
	\label{fig:framework-concept}
\end{figure}

This process is visualized in \autoref{fig:framework-concept}.
On this level of abstraction, we can already postulate:
Usage tracking requires oversight over all data processing. Therefore, integrating it into the analysis tool itself is ideal. That matches our concept that these tools be built with inverse transparency by design.
But it seems as if the tasks of storing monitored usages and making them transparent could be extracted and shared between multiple tools.
In that case, we could reduce development effort of tool developers and potentially increase security, as only one database would need to be protected.
For data owners, having a single tool to watch over how their data are used has the potential to reduce mental load and may therefore be preferable, too.

\subsection{Implications for Software Design}
\label{sec:concept-implications-software-design}

In order to make all data usages transparent, we need to not only track all occurring usages, but also prevent circumvention of the framework logging. Therefore, it is important to ensure that data never leave controlled environments. Accessing them may only be allowed through tools that provide inverse transparency, thereby ensuring that usages are logged~\cite{weitzner2008information}.
This is why we envision inverse transparency by design: No software tool should be distributed without enabling this transparency.
In our concept, this means that tools need to, on a conceptual level, include a \emph{Monitor} that tracks data processing.
We can then imagine companies running their own, private log store providing a standardized API for adding and retrieving log entries.
Tools they license or deploy are then required to integrate into the company inverse transparency infrastructure, sending the logged data usages to their private \emph{Safekeeper} API.

At first glance, this seems reasonable. Yet, when we look closer, we find that data are usually not ``sent'' to data consumers, but continuously move between (sub) systems, are aggregated, copied, converted, or moved. A ``data usage'' may for example just be a software tool starting up---simply to show a data-driven start page, data accesses can be necessary. Therefore, we have to consider what a data usage is and how software developers can ensure that it is logged.
For most cases, our concept of developing with inverse transparency by design deals exactly with this problem: Instead of trying to retrofit usage logging to existing software, we expect developers to already recognize the interests of data owners in the design phase.
In some cases, though, building with inverse transparency in mind may require a paradigm shift.
Specifically, consider a scenario that we refer to as \emph{ambient usage} of data.
We illustrate this with a typical home page of an analytics tool.
\begin{figure}[htbp]
	\centering
	\includegraphics[width=.9\linewidth]{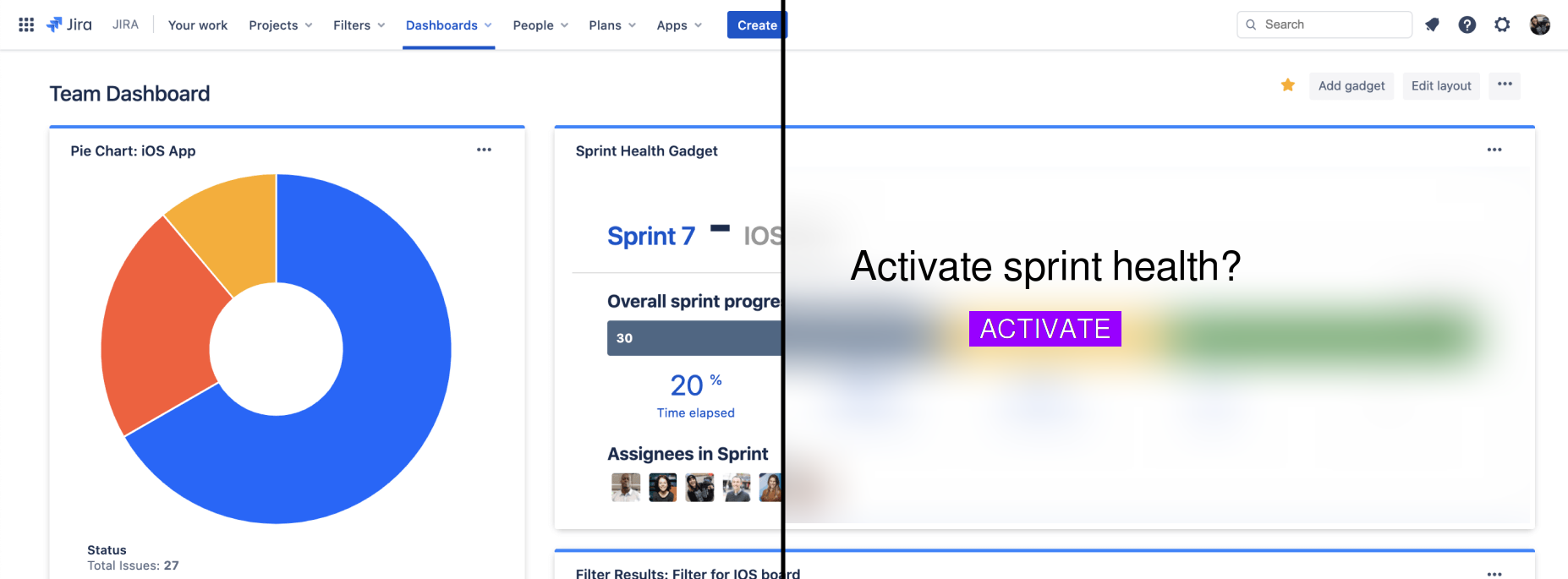}
	\caption{
		Example of how interaction paradigms may change as a consequence of inverse transparency, illustrated at the example of Jira Software (image source: Atlassian; modifications: the authors).
		On the left side, the status quo in Jira: analyses are run and insights presented automatically when the tool is opened, a pattern we refer to as \emph{ambient usage} of data.
		The right side illustrates how inverse transparency may transform this paradigm: the insights are hidden and need to be explicitly activated, signaling intent by the data consumer.
	}
	\Description{
		A dashboard of Jira Software is shown, visually separated in the middle by a black line. On the left, the dashboard is unchanged, displaying various analyses. On the right, it is blurred, with a text ``Activate sprint health?'' and a button ``ACTIVATE'' below it.
	}
	\label{fig:jira-dashboard-ambient-usage}
\end{figure}
On load, the tool presents multiple widgets that represent analyses, ready for the data consumer to view (see \autoref{fig:jira-dashboard-ambient-usage}, left side).
If we now consider inverse transparency, merely opening the tool already would result in a multitude of recorded data usages, even if none of these widgets are viewed by the data consumer.
Data owners may be unnecessarily worried about data usages, and data consumers may need to justify their presumed interest.
Conversely, this plausible deniability would make it nigh impossible to differentiate if the data consumer actually made use of the analyses or not.
A potential consequence of this may be a fundamental shift in interaction paradigms. Instead of presenting all possible data up-front for data consumers to view, the tool may present blurred windows with a button to explicitly ``show'' the respective analysis (see \autoref{fig:jira-dashboard-ambient-usage}, right side).
Thereby, the consumer specifically expresses their intent to access the data shown, with the corresponding log entry being written in the background.
We can conclude that, in the long term, inverse transparency may result in software being implemented with a more mindful approach to data usage, moving away from the currently common arbitrary data processing.

\subsection{Usage Log Integrity}

After considering data flow tracking, we shift our focus to the usage logs and the \emph{Safekeeper}.
The integrity (completeness and correctness) of the stored logs is a central requirement for inverse transparency to function.
Only then can we achieve accountability of data consumers.
Considering completeness, we have deliberated how building software with inverse transparency by design ensures that any data usage can be detected.
Yet, even if our approach is adopted immediately, there will be a transitional period, which means we need to contemplate how to enable usage logging for existing systems that have not been built this way yet.
Indeed, this exact issue has been worked on extensively in the context of usage control. Concretely, the research on distributed usage control~\cite[see, e.g.,][]{pretschner2006distributed, pretschner2014achieving, kelbert2018data} tackles this problem and can be seen as a complement to ours. For example, \citeauthor{loerscher2012data} showed how to implement the necessary data flow tracking for the Thunderbird mail client~\cite{loerscher2012data}.
Recently, application sandboxing has been proposed to achieve similar goals without requiring changes to the monitored tool~\cite{kraska2019schengendb}.
We believe that a combination of distributed usage control and our approach is sufficient to cover all potentially occurring data usages, providing reasonable completeness of the logs.

\begin{figure}[htbp]
	\centering
	\includegraphics[width=\linewidth]{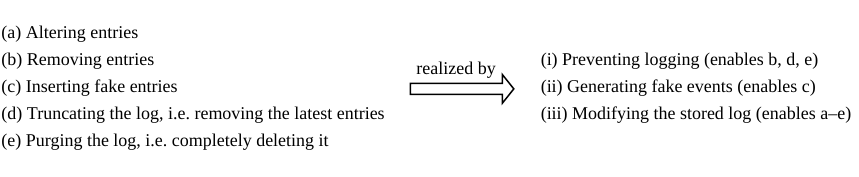}
	\caption{Conceptual attacks on the usage log~\cite[following][]{ahmadvand2017distributed} and their realization.}
	\Description{Two lists are shown. On the left, a list of conceptual attacks, labeled a through e. They are, in order, altering entries, removing entries, inserting fake entries, truncating the log, and purging the log. An arrow points to the second list labeled ``realized by''. The second list is the list of realizations, labeled i through iii. They are, in order, preventing logging (enables b, d, e), generating fake events (enables c), and modifying the stored log (enables all a–e).}
	\label{fig:conceptual-attacks-and-realizations}
\end{figure}

Therefore, we now consider conceptual attacks on our stored usage log.
We find five abstract approaches to attack the log integrity that can be realized through three main attack vectors (see \autoref{fig:conceptual-attacks-and-realizations}).

In this paper, we do not present detailed technical solutions for these attacks. Yet, we should consider if any of these attacks may make our idea conceptually infeasible before we continue.
And, while we cannot conclusively judge the applicability of these attacks on this level of abstraction, we do find sensible solutions for them that suggest they might be manageable.
For (i), a reasonable expectation would be DDoS attacks. In literature, we find a number of papers and books describing how to address and prevent such attacks~\cite[e.g.,][]{bhattacharyya2019ddos}.
For (ii), we find two attacks reasonable to expect: constructing a realistic-looking fake log event or replaying past events.
Again, existing literature presents various ways to approach and prevent such attacks.
On the one hand, to prevent attackers from simply faking log events, applications should sign requests to ensure their authenticity~\cite[see][]{delfs2007public}.
Simple strategies to mitigate replay attacks, on the other hand, are also well researched and seem reasonably easy to implement~\cite[e.g.,][]{aura1997strategies}.
Accordingly, we consider (i) and (ii) to be mainly engineering challenges for the specific instantiation, yet ultimately solvable.

Attack variant (iii) meanwhile is much broader and can encompass various strategies, including hardware-based attacks such as removing a hard disk containing the logs from a server rack.
The field of secure logging, which is concerned with mitigating such attacks, is vast.
That means we can only roughly sketch potential avenues that exist to protect against various kinds of malicious log modifications.
For example, in our research we found existing solutions based on trusted computing modules~\cite[see, e.g.,][]{accorsi2010bbox, priebe2018enclavedb, lee2020secure}, cryptographically secured protocols such as forward-secure authentication~\cite[see, e.g.,][]{holt2006logcrypt, ma2009new, ahmadvand2017distributed}, or distributed ledger technology, i.e. blockchain~\cite[see, e.g.,][]{ge2019permission, schaefer2019transparent, zieglmeier2023decentralized}.
Recent challenges seem to include heterogeneity of hardware~\cite{lee2020secure} and GDPR compliance, especially considering the right to erasure~\cite{zieglmeier2021gdpr}.
Yet, these are robust approaches that claim to guarantee integrity and even confidentiality of the stored logs against reasonably powerful adversaries.
That means, while there exist some open research questions, we consider the problem of log integrity to not be in fundamental conflict with our concept.

Of course, this glosses over one important caveat: malicious data consumers.
As soon as data are provided to users, even within a monitored environment, our control over it ends.
How do we deal with data export functionality or with tools that store data on disk, meaning these data may be accessed by malicious data consumers without utilizing a monitored tool?
In short, this may not be necessary at all.
While it is still common today to allow users access to the underlying data and files, businesses are quickly moving towards a future of cloud software, provided only as a ``service''~\cite{gurses2018privacy}.
There, data are merely an enabler of features, with tools moving away from the idea of ``files'' as containers for data that are manually handled by users.
The corresponding loss of control for users has become so significant that legislation such as the GDPR specifically includes rights to data export and portability for users' own data.
While this trend to software-as-a-service arguably reduces users' data sovereignty, it allows us to consider the problems of data exports or files stored outside our control to be solved.
And while it naturally is, in any setting, nigh impossible to prevent data consumers from simply taking a picture of the screen~\cite{pretschner2008usage} or even just memorizing its contents~\cite{pretschner2013distributed}, we argue that this does not significantly reduce the value of the provided transparency.
On the one hand, measures to control this would be questionable in their effectiveness and highly invasive (e.g., eye tracking), making them unreasonable to consider.
On the other hand, large-scale copying or exporting of data is not feasible this way, making it only a theoretical issue.

\subsection{Trustworthiness}
\label{sec:concept-trustworthiness}

From a technocratic worldview, it might seem as though the only relevant factor to concern us to enable inverse transparency would be the log integrity discussed above. After all, as long as everything is logged in a tamper-proof way, data owners should be satisfied.
This is not the case, though.
To enable accountability, we need data owners to accept and trust the provided transparency, and, as a consequence, make use of it.
Above, we have deliberated how the conceptual tasks of \emph{Monitor} and \emph{Safekeeper} can enable inverse transparency by protecting log integrity.
In the following, we consider the third part, the \emph{Display}, representing the user interface making stored usages transparent to individuals.

Our goal is to enable data owners to make use of the transparency provided to them, enabling their data sovereignty.
We find two potential obstacles: individuals not being able to operate the provided transparency interface or understand its contents (usability), and not being able to trust it (trustworthiness).
While it is clear that usability is important, user trust in technology is equally relevant, considering that it influences users' intention to use, adoption, and continued use of a tool~\cite{riegelsberger2005mechanics, sollner2012understanding}.
We have already covered an important prerequisite for data owners to trust the provided transparency---ensuring the integrity of the log---but we should not underestimate the influence of the design of the user interface on their trust.
The way information is presented, framed, and the human factors that surround a system's implementation or rollout are of high relevance for user trust~\cite{fisher2020does, zieglmeier2021designing}.
To illustrate this: If a system is built to be cryptographically secure, but users are not informed of this fact or do not trust the messenger, this technical fact alone will do little to improve their trust in the system.

Following \citeauthor{zieglmeier2021designing}, we therefore consider the three trust dimensions of \emph{purpose} (the intended use of a system), \emph{process} (how it operates), and \emph{performance} (how well it solves its tasks) that are relevant for user interface design~\cite[pp.~2--3]{zieglmeier2021designing}.
As the operation of the system---its process and performance---depend on the concrete instantiation, we therefore focus on its perceived \emph{purpose} here, a facet that could threaten our idea on a conceptual level already.
In short, the purpose dimension of trust reflects ``the impression of the designer’s intentions that users get from interacting with the system''~\cite[p.~2]{zieglmeier2021designing}.
It becomes clear that the intended use of the transparency system is core to its trustworthiness. When introducing an inverse transparency infrastructure, a company needs to ensure that employees trust the system to serve their goals and not those of the company.
If this is not addressed, the system may in the worst case not be accepted by employees at all, therefore rendering it meaningless.
Accordingly, we need to consider whether there are ways to ensure, on a conceptual level, that the provided transparency is experienced as trustworthy by individuals.
When researching approaches to this problem, we find multiple promising solutions.
For example, to improve trust, a reputable or well-known (third) party can be made responsible for the development and operation of the transparency-enabling systems or certify their correctness, thereby targeting the trust antecedents of reputation and familiarity~\cite{hoff2015trust}.
If available, this could be the company-internal workers' council, or alternatively an external workers' rights organization.
Should that not be possible, the use of open-source software or code audits can reduce the necessity of interpersonal trust~\cite{garcia2005trust, alarcon2020trust}.
We find both approaches to be reasonable and realistic to implement.
Therefore, we conclude that the trustworthiness of the transparency system is not in fundamental conflict with our concept.

\section{Study A: Software Developer Perspective}
\label{sec:case-study-developer-perspective}

Our concept of software being built with inverse transparency by design necessitates a behavioral and mindset change in software developers.
Therefore, it is important to consider their perspective to understand potential conceptual issues early. If developers find the concept infeasible to implement in practice, we need to address their issues first before we can continue.
Furthermore, we strive to learn about the implications of our concept for software design. As with any restriction on how software should be developed, this may hinder innovative features, or enable completely new solutions to problems not imagined before.
We therefore present a preliminary study of the developer perspective. It is designed to answer the following research questions:

\newcommand{\devstudyRQitbdManifestsInArtifact}{How does developing people analytics with inverse transparency by design manifest itself in their architecture?}

\newcommand{\devStudyRQitbdJudgedByDevs}{How is the approach of inverse transparency by design judged by developers?}

\begin{enumerate}
	\renewcommand*\theenumi{A.\arabic{enumi}}

	\item \devstudyRQitbdManifestsInArtifact
	\label{rq:itbd-manifests-in-artifact}

	\item \devStudyRQitbdJudgedByDevs
	\label{rq:itbd-judged-by-devs}
\end{enumerate}

\subsection{Study Approach}

We opted for an artificial setting designed to closely mirror a real-world software development use case.
To that end, we created a university practicum spanning over three months with 12 computer science master's students as participants.
By fully controlling the setting, we were able to choose the concrete tasks worked on and allow participants appropriate time and resources to implement software according to our principles, establishing causality~\cite{sjoberg2005survey}.
Working with students, meanwhile, allowed us to best test our initial hypotheses~\cite{tichy2000hints}.
Thereby, we strove to exclude entrenched mental models~\cite{uitdewilligen2013mental} and development culture~\cite{ambler2008agile} as confounding factors.
For our case of applying a new approach for the first time, students have been shown to perform comparably to professionals~\cite{salman2015students}.

Participants were tasked to develop software in agile development teams.
Four groups of developers were formed, considering individuals' skill level and technology preferences.
The team members had never worked together before, removing any potential influence of an existing development culture.
In the first two months, three teams were tasked to develop people analytics with inverse transparency by design (covering the \emph{Monitor} in our concept), while the fourth team implemented and improved the necessary auxiliary tools for the concept (\emph{Safekeeper} and \emph{Display}) based on feedback from the other teams.
In the final month, participants used the developed analysis tools to analyze their own data collected in the months prior. This allowed them to experience the analyses both from the perspective of a developer and a user.

The development tasks for the analytics teams were derived from a set of use cases that we developed with our industry partner.
Teams independently chose their concrete task from the set based on technical skills and interests.
The provided use cases covered potential data sources as well as relevant insights.
Thereby, we made sure that the developed analytics were grounded in practice.
For a scientific grounding, meanwhile, teams were tasked to read relevant academic literature as part of their development and link those insights to their implementation decisions.

To foster critical reflection and deliberation among the developers in our study, we instructed them to integrate the ethical deliberation for agile processes (EDAP)~\cite{zuber2020ethical} into their work process.
EDAP is a methodology to interweave ethical deliberation with traditional agile development projects. The goal is to introduce normative deliberation to developers about which technical direction is preferable for their software~\cite[pp.~11--12]{zuber2020ethical}.
We found the EDAP schema to be an effective tool to foster reflections and deliberations from participants about their software development projects.
The development teams were instructed how to perform an analysis following the EDAP scheme. They were tasked to update their report bi-weekly during the implementation phase. After the last week of development, the deliberation report was finalized and submitted.
In addition, after the conclusion of the study, teams submitted their code as well as a longer written analysis reflecting on the impact of inverse transparency on their implementation.

\subsection{RQ~\ref*{rq:itbd-manifests-in-artifact}: \devstudyRQitbdManifestsInArtifact}

The feasibility and implications of inverse transparency by design as a development approach can be seen in the architecture of people analytics developed according to the principle.
Therefore, we investigate each of the three artifacts implemented by the analytics development teams in our study.
To preface, we find that no team encountered fundamental issues with the process and all could implement their envisioned projects.

\subsubsection{Team G: Version Control Software Analysis Tool}

\begin{figure}[htbp]
	\centering
	\includegraphics[width=.85\linewidth]{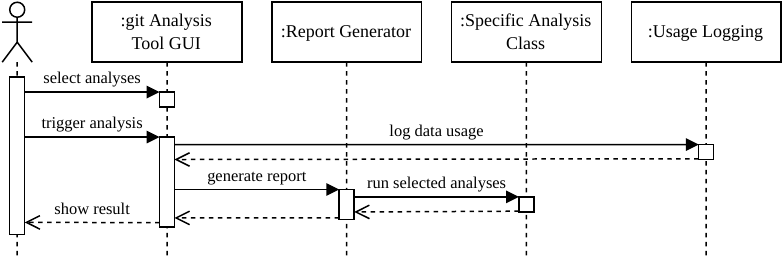}
	\caption{
		Sequence diagram of how team G realized inverse transparency. Data consumers have to explicitly select analyses before any data are processed. This data usage is logged once, before the report is generated.
	}
	\Description{A sequence diagram is shown that graphically represents the process as described in the text.}
	\label{fig:study-devs:sequence-diagram-git}
\end{figure}

The first team implemented a history analysis for Git\footnote{\url{https://git-scm.com}} commits in a standalone application.
One example for an analysis is the \emph{commit hours} overview. It summarizes for each developer the number of commits per hour of the day.
The analyses are implemented as tabs that the data consumer can switch between.
To implement inverse transparency, they include an additional screen in their tool that may not be necessary otherwise: a selection screen for the requested analysis.
It manifests a simplistic solution for the issue of \emph{ambient usage} in the context of inverse transparency that we deliberated above (see \autoref{sec:concept-implications-software-design}).
Only those analyses that the data consumer explicitly selects are loaded, with a usage log immediately created for every selected one (see \autoref{fig:study-devs:sequence-diagram-git}).
After the analysis report has been created, the tool logs no additional accesses even if it is opened again, which is interesting albeit inconsequential.
We consider this the most basic form of implementing inverse transparency: the interaction paradigm and provided features do not have to change, instead the tool adds a step to the process before starting the data processing and just records any potential data usage immediately.

Team G worked together with team S on their implementation of inverse transparency. They implemented a shared library that was then integrated by both into their code. Hereby, we could already see potential synergies when software is implemented with inverse transparency by design:
Basic functionality does not have to be reimplemented for every tool and may instead be shared, minimizing the development overhead.

\subsubsection{Team J: Issue Tracking Software Analysis Plugins}

\begin{figure}[htbp]
	\centering
	\includegraphics[width=.85\linewidth]{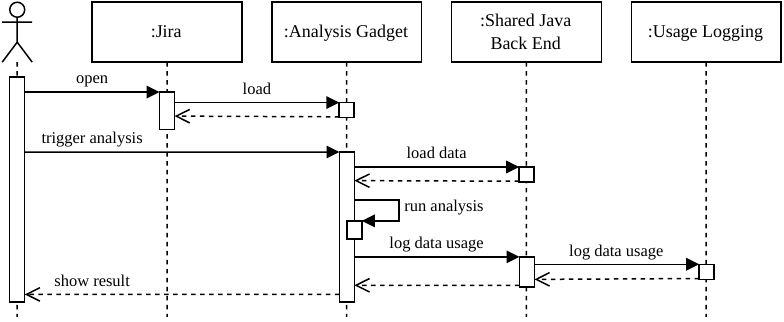}
	\caption{
		Sequence diagram of how team J realized inverse transparency. A shared back end component serves as the single point for accessing data and logging usages. Contrary to team S' instantiation, the usage is logged based on the analysis \emph{result}, not the consumer's \emph{request}.
	}
	\Description{A sequence diagram is shown that graphically represents the process as described in the text.}
	\label{fig:study-devs:sequence-diagram-jira}
\end{figure}

Team J implemented four analysis gadgets for Jira Software,\footnote{\url{https://www.atlassian.com/software/jira}} distributed as one plugin.
One example is the \emph{supporter analysis}. It ranks team members by who performed most code reviews.
Data consumers can then choose any of the gadgets independently to be added to their main application dashboard, where they have to be explicitly activated to run.
Considering its architecture, the artifact implements inverse transparency in a notable fashion---a single module provides both data retrieval and usage logging functionality, allowing the various plugins to share this code.
Thereby, developers do not need to consider inverse transparency for every newly created analysis.
In the front end code, this manifests as a single function invocation that logs the data usage via the shared back end (see \autoref{fig:study-devs:sequence-diagram-jira}).
After it completes, the tool continues its operation. On failure, it shows an error message, ensuring that the logging successfully completes before data are presented to the data consumer.

\subsubsection{Team S: Chat Analysis Tool}

Team S implemented a standalone analysis tool for Slack\footnote{\url{https://slack.com/}} workspaces.
For example, one analysis their tool can perform is the \emph{network analysis}. It creates a social network graph based on who is ``mentioned'' by whom in their messages.
Various analyses such as this can be chosen on the main screen and then independently triggered with a button.
\begin{figure}[htbp]
	\centering
	\includegraphics[width=.85\linewidth]{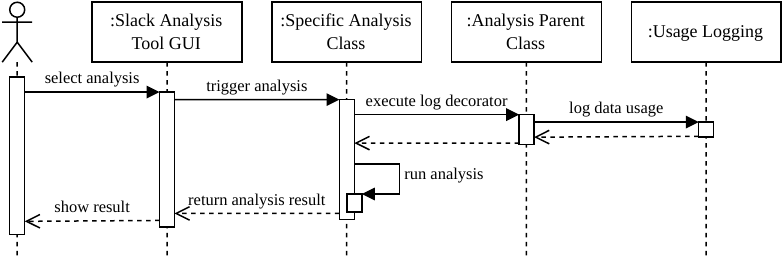}
	\caption{
		Sequence diagram of how team S realized inverse transparency. The log decorator logs usages after the analysis is triggered but before it is run. The architecture prevents new analyses to be implemented without usage logging.
	}
	\Description{A sequence diagram is shown that graphically represents the process as described in the text.}
	\label{fig:study-devs:sequence-diagram-slack}
\end{figure}
For team S, thinking with inverse transparency by design meant they built their whole architecture around the concept.
First, their program code is structured as such that no new analysis function can be added without inverse transparency.
If occurring data usages are not logged before an analysis is run, the code triggers an error that causes the application to close.
Second, to implement this logging, they provide a code tag (``decorator'') that can be added to new analysis functions (see \autoref{fig:study-devs:sequence-diagram-slack} and \autoref{lst:analysis-decorator})
\begin{lstlisting}[
float=htbp,
frame=tb,
caption={Simplified source code of team S. Their architecture is structured so that, to provide inverse transparency, they only need to decorate the function performing the data analysis with \texttt{@Analysis.log\_data\_usage}. This decorator ensures that the data usage is logged whenever the analysis is requested.},
label=lst:analysis-decorator,
basicstyle=\ttfamily,
language=Python]
class NetworkAnalysis(Analysis):

    @Analysis.log_data_usage
    def perform_analysis(self) -> Result:
        # ...
\end{lstlisting}
This tag then handles the usage logging automatically.
That means that developers adding new analyses only need to add a single line to their program code to integrate inverse transparency.

To run an analysis, data consumers select it from the application's main screen. Only when an analysis is explicitly triggered, the data usage is logged, after which the results are shown.

\subsection{RQ~\ref*{rq:itbd-judged-by-devs}: \devStudyRQitbdJudgedByDevs}

Second, we consider developers' perspectives on the idea of embedding inverse transparency into people analytics.
To that end, we analyze the ethical deliberations and reports of the development teams.
We find that, in general, the teams show neutral or positive sentiment towards developing with inverse transparency by design.

In the following, we start by investigating the deliberations of all three analytics development teams combined.
We reference the source of each quotation by referring to the respective team's letter in parentheses, e.g. (J) for team J.
Then, we shift our focus to the fourth team, tasked with developing the auxiliary tools for inverse transparency, notably the user-facing \emph{Display}.

\subsubsection{Analytics Developers}

To start with, the developers recognize the value of the various implemented analytics, noting that they can be ``desirable for both a comfortable and effective work environment'' (J).
They highlight that the tools can provide ``a more objective view'' and help ``verify [ones] hypotheses about a team'' (G).
This shows that our participants do not fundamentally oppose the idea of implementing people analytics.
Yet, they also point to the risk of such tools, as their insights could be considered ``privacy invasive'' (J), ``flawed'', or ``biased'' (G).
Furthermore, they could ``become a [...] self-fulfilling truth'' if relied on unquestioned (S).

To counter the risks to some degree, ``access control and inverse transparency with regards to data access is desirable'' (J).
Integrating inverse transparency by design, they argue, ``increases transparency'' and ``gives control to the owners of data'' (J).
One important concern with people analytics are indirect negative effects, such as users adapting their behavior to conform to expectations.
This issue is explicitly acknowledged by team S, and they argue that ``inverse transparency [...] should help reduce [this] pressure to conform for the users.'' (S).
One potential reason for this is that inverse transparency is judged to be capable to ``avoid misuse of [the tool]'' (J),
which is ``likely to reassure the user of the safety of their data'' (J)
and ``[face] the concerns of data owners'' (G).

Regarding the technical realization with inverse transparency by design, the developers note no issues.
To the contrary, team S argues that they ``demonstrated [...] that inverse transparency is a viable concept which can, at least from a technical standpoint, work in practice'' (S).
This matches our findings in the analysis of the implemented artifacts.

Providing additional protection through inverse transparency is judged to be valuable and technically feasible.
Yet, this may also have unexpected side effects on development decisions.
Team S worked on analyses that can be considered, in parts, relatively invasive.
They acknowledge this in their EDAP report, noting a conflict between ``stakeholder evaluation of / knowledge about employees'' versus ``privacy of workers, freedom of surveillance''.
As one solution for this conflict though, they note that ``inverse transparency [...] ensures that the employee is at all times informed of the extent of the analysis''~(S).
This hints at an unexpected aspect of developing with inverse transparency by design: It might actually serve as an enabler of features commonly considered privacy-invasive or delicate, as it could to an extent counteract their negative consequences.
We deliberate this point further in \autoref{sec:discussion}.

\subsubsection{Inverse Transparency Tool Developers}
\label{sec:dev-study-deliberation-t}

Second, we analyze the deliberations of the fourth team, tasked with implementing the auxiliary inverse transparency tools.
Most interesting is their work on the user-facing \emph{Display}, which makes data usages visible to data owners.
For them, this interface can be seen as the manifestation of inverse transparency.

There is always a manipulative element when designing a user interface.
Choosing if and how to display certain information naturally influences how it is perceived.
The team recognizes that concern.
They note that, on the one hand, data owners of course need guidance ``on how [the data] should be interpreted.''
Yet, providing such guidance by, e.g., coloring specific values has a risk of ``arbitrarily highlight[ing] certain usage scenarios and not others.''
This could be especially critical if data consumers are unaware of being singled out.
Therefore, they propose to assist data consumers in understanding how their logged data usages will be displayed.
As an example, one could ``provide documentation on what presentation [they] can expect as a result of a certain action.''

The team discusses such issues under the notion of \emph{fairness} towards the data consumer.
They warn that even ``facts [...] presented by the system [...] alone may convey the wrong impression of a data consumer which could violate the principle of fairness.''
Therefore, they argue for allowing data consumers to provide explanations for data usages ``to justify their actions.''
Yet, they also recognize that data consumers may have an interest to manipulate and therefore ``caution has to be taken.''
We find this a very important issue that is critical to deliberate.
Providing inverse transparency necessarily also leads to increased transparency over data consumers' actions.
Including their perspective in the design of the transparency tools is therefore, in our view, necessary.

\subsection{Conclusions}

In our study of the developer perspective, we find that developing people analytics with inverse transparency by design is technically feasible.
This can be seen from the artifacts as well as developers' reflections.
The teams furthermore consider inverse transparency capable to counter some of the risks of people analytics.
Considering their implementations of inverse transparency, team J's and team S' architectures are completely built around the usage logging.
This prevents accidental circumvention of the logging and facilitates the integration of new analyses with inverse transparency.
Refactoring an existing codebase this way retroactively, meanwhile, could mean significant additional effort and technical risks~\cite[see, e.g.,][]{sharma2015challenges}.
Additionally, all implemented artifacts require data consumers to explicitly select and trigger analyses before the results are shown.
Thereby, they counteract the issue of \emph{ambient usage} of data that we identify in \autoref{sec:concept-implications-software-design}, as this step prevents unintended data usage.

These findings match our expectation that building with inverse transparency by design has noticeable influence on the architecture and interaction paradigm of people analytics.
Accordingly, they reaffirm our approach compared to retrofitting transparency, and show its feasibility.

\section{Study B: User Perspective}

After considering the developer perspective, we shift our focus to the users in a second preliminary study.
Inverse transparency can only unfold its full potential if users find it helpful and make use of it to enable accountability.
Therefore, our second study is focused on assessing the effect that inverse transparency by design can have on employees and how they might experience it.
Our research questions are:

\newcommand{\usrStudyRQunifiedDashboard}{How is a unified inverse transparency dashboard judged by participants?}

\begin{enumerate}
	\renewcommand*\theenumi{B.\arabic{enumi}}

	\item \usrStudyRQunifiedDashboard
	\label{rq:unified-dashboard}

	\item Do data owners find inverse transparency beneficial and feel empowered by it?
	\label{rq:helpful}

	\item Do participants consider inverse transparency to be capable to influence data consumers' data usage behavior?
	\label{rq:influences-consumers}

	\item Is inverse transparency considered valuable in general by participants?
	\label{rq:valuable}

\end{enumerate}

\subsection{Study Approach}

We conducted a laboratory study with students, which was specifically designed for our research objectives.
Inverse transparency represents a fundamental change in the working environment and interactions of employees.
Introducing it to an existing workplace could have triggered confounding change management issues, most notably individual's resistance to change~\cite{bovey2001resistance}, which can limit the success of change initiatives~\cite{kuhlman2021will}.
By working with students and fully controlling the study environment, we could ensure that inverse transparency was actually experienced and used over an extended period of time~\cite{katok2018designing}.
Furthermore, having removed external influences of an existing workplace, such as schedule pressure~\cite{nan2009impact}, means that we can cleanly attribute any observed effects to inverse transparency instead of potential confounders~\cite{fiske2005laboratory}.
Accordingly, we created a university practicum running over three months for our study.
To increase the representativeness of the study environment, we emulated the real-world use case of a remote software development workplace as closely as possible.
That means having small teams of developers build software in agile teams, each lead by a team lead. Work items and their status were tracked in the issue tracking software Jira Software, messages were exchanged over the business messenger Slack, and code was versioned with the distributed version control software Git.
Teams only interacted through digital collaboration tools, representing an all-remote configuration.

\begin{figure}[htbp]
	\centering
	\includegraphics[width=0.55\linewidth]{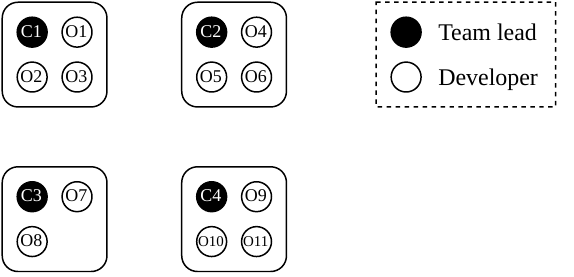}
	\caption{Participant organization in our study ($n = 15$). Three teams of four and one of three were formed. Each team had one team lead, serving as a data consumer (C1--C4), with the rest of the team working as developers, representing data owners (O1--O11).}
	\Description{An abstract overview over the participants and their teams is shown. Four boxes contain circles that represent participants. Three boxes contain four, one contains three. Each box has one black circle, representing the team lead. They are labeled C1 through C4. Additionally, they have three, respectively two, white circles for the developers. These are labeled O1 through O11.}
	\label{fig:study-overview_pit21-teams}
\end{figure}

In total, 15 master's or final year bachelor's students in computer science were tasked to conduct work in the setting described above.
Working with computer science students specifically allowed us to realistically model the software development use case.
They were split up in four groups of four students each, one with three (see \autoref{fig:study-overview_pit21-teams}). Each group decided on a team lead to guide the development process and serve as the data consumer.
We explicitly gave team leads the task to conduct data-driven management, utilizing analysis tools built with inverse transparency by design to analyze their team members' data and write reports on their performance.
Due to the all-remote configuration, team leads had to rely on these analytics to get a full picture of their teams.
The other team members worked as developers, representing the data owners in our concept.
They were instructed to specifically utilize those data-driven collaboration tools that their team leads' analysis tools were compatible with. That means they worked with Jira, Slack and Git, as described above.
The data collected in these tools were then made available to their respective team leads to analyze via people analytics built with inverse transparency by design.

We employ a mixed-method evaluation design.
During the process, regular written self-reflections were submitted.
In addition, participants answered two questionnaires after the study, one of which covers their experience with inverse transparency over the course of the study, the other their opinion of inverse transparency in general (the questions are listed in \hyperref[sec:appendix-questionnaires]{Appendix~\ref*{sec:appendix-questionnaires}}).
Our findings summarize the results from all of these evaluations.

\subsection{Workflow and Utilized Tools}

\begin{figure}[htbp]
	\centering
	\includegraphics[width=.85\linewidth]{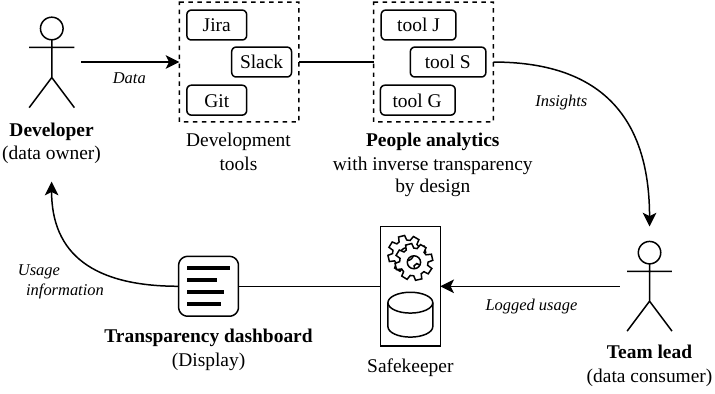}
	\caption{The workflow and utilized tools in study B. Developers worked with development tools that collect data on their work and collaboration. These data were then analyzed by the team leads with the people analytics created in study A. The tools logged their data usage, storing this data in the \emph{Safekeeper}. The usage information was then made available to developers via the transparency dashboard.}
	\Description{A cycle is shown. Top left, a figurine labeled 'developer'. From it, an arrow labeled 'data' into a box 'development tools'. From that, an arrow 'insights' through a box 'people analytics with inverse transparency by design' to a figurine labeled 'team lead' in the bottom right. From that, an arrow 'logged usage' to 'Safekeeper' left of it. Finally, from that, an arrow 'usage information' through a box 'transparency dashboard (Display)' back to the figurine 'developer' in the top left.}
	\label{fig:study-users_workflow}
\end{figure}

The teams in study B worked with people analytics that were built by the developers in study~A with inverse transparency by design.
The employed workflow, visualized in \autoref{fig:study-users_workflow}, was an instantiation of our concept (see \autoref{fig:framework-concept} on page~\pageref{fig:framework-concept}), with developers representing data owners and team leads representing data consumers.
Developers used the tools Jira, Slack, and Git for their work and collaboration.
The data collected in these tools were then accessed by the team leads via the people analytics created in study B by team J, team S, and team G.
As these analytics were built with inverse transparency by design, any data usage was logged and stored in a database (the \emph{Safekeeper}) to be made available to the data owners.
To access usage information relating to their data, developers were given access to a transparency dashboard (the \emph{Display}) that provides a direct view into the tracked usage logs.
This dashboard was developed by the fourth team in study A as part of the auxiliary transparency tools, in addition to the \emph{Safekeeper} that stored usage data, and a single sign-on server.
Collectively, we refer to these tools as the \emph{inverse transparency toolchain}~\cite{zieglmeier2023inverse}.

The purpose of the transparency dashboard is to make usage information available to data owners.
It is a standalone tool that unifies all logged usage data relating to an individual, independently of the concrete analytics used.
All study participants got accounts for the inverse transparency toolchain that they could use to log into the transparency dashboard.
To illustrate our following elaborations on the dashboard, find a screenshot of it in \hyperref[sec:appendix-screenshot-clotilde]{Appendix~\ref*{sec:appendix-screenshot-clotilde}}.
After logging in, users find two views into the data: an overview area and a detailed table with individual entries.
The overview area is meant to give a quick view how one's data were used in the past days.
To find out exactly what was accessed, by whom, and when, the table and its filtering functionality can be used.

\subsection{RQ~\ref*{rq:unified-dashboard}: \usrStudyRQunifiedDashboard}

To begin with, we evaluate how participants judge having one unified inverse transparency dashboard.
Alternatively, individual dashboards could be integrated into each tool.
We theorize in \autoref{sec:concept} that having a single point summarizing all usage information could be preferable.

To exclude potential usability issues that may have impacted participants' experience, we used the commonly applied system usability scale (SUS)~\cite{brooke1986system} (Q1--Q10).
The aggregated SUS score of \emph{81.67} indicates that no fundamental usability issues arose during use~\cite[see][]{bangor2008empirical}, supporting our qualitative results.
Then, we asked participants to formulate freely if they enjoyed using the dashboard and what could have been improved (Q11 \& Q12).
Both questions aim to uncover benefits and issues of our approach of providing a unified dashboard.

Participants unanimously expressed positive sentiment about the unified transparency dashboard, describing it as ``useful'' (O11, C3), ``simple and easy to use''~(O2) and noting that it ``gets things done fast'' (O3).
Referring to its concrete value, data owners note that ``being able to see [...] data access patterns from managers was very interesting'' (O8) and that ``it is very useful to have an overview of the data usage'' (O11).
This ability to detect usage patterns, by getting an overview of data usages, is facilitated by all data being presented in one unified dashboard.
For data consumers, meanwhile, this may provide a different benefit.
One notes that the unified dashboard was ``a useful tool when trying to determine whether my data accesses were logged properly'' (C3).
As we discuss in our study of the developer perspective (see \autoref{sec:dev-study-deliberation-t}), data consumers may want to verify that their data usages are reported correctly.
Having a single dashboard that unifies all collected data can make this easier.

When asked what could be improved, participants mentioned a need for more visual summaries of the recorded accesses (O1, O2, O11, C4) and tool tips or explanations for what is displayed~(O4, O7).
Providing visual summaries can be especially useful when unifying data from multiple sources into one dashboard, as this may help uncover suspicious usage patterns.
Two responses support this point by referring to our approach of a unified dashboard directly, with both stressing its value.
In fact, these participants wished for even more of the utilized tools to be integrated into the transparency infrastructure: ``Every module [...] should integrated [sic!] into [it]'' (O3), as this ``would be very interesting'' (O8).
This confirms our vision; we are convinced that inverse transparency can unfold its full potential only if it is implemented as a foundational paradigm underlying all people analytics, not just a selected few.
Our results suggest that a unified dashboard can facilitate this.

\subsection{RQ~\ref*{rq:helpful}: Do data owners find inverse transparency beneficial and feel empowered by it?}

Inverse transparency as a concept mainly targets data owners---those who provide data for others to use.
As described above, the data consumers in our study were explicitly instructed to utilize tools built with inverse transparency by design to analyze data collected from Jira Software, Slack, and Git. These data were generated by data owners as side effects of their work. The data usages were then recorded and made available to the data owners on a dashboard to provide them with inverse transparency.

\begin{figure}[htbp]
	\centering
	\includegraphics[width=\linewidth]{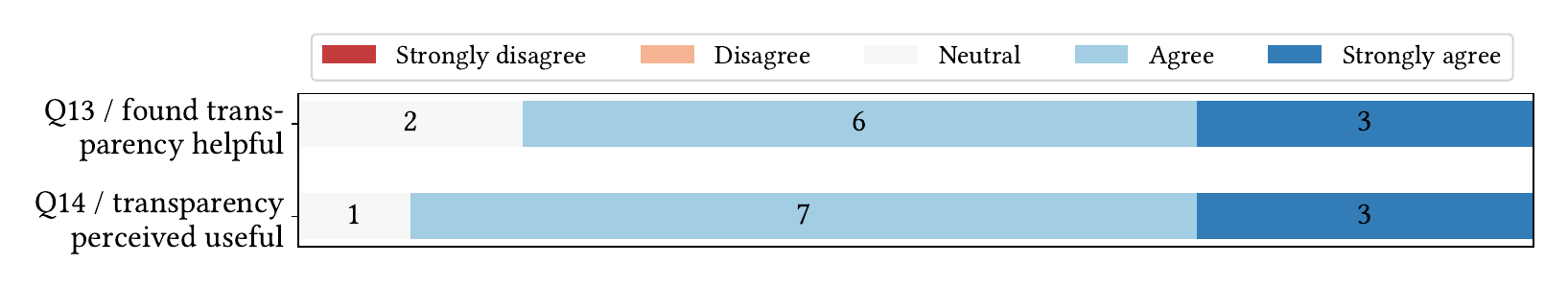}
	\caption{Results of the questions for data owners ($n = 11$) about how helpful and useful they perceived inverse transparency to be. They were asked to express their agreement on a five-point Likert scale (X-axis), from 1 for ``strongly disagree'' to 5 for ``strongly agree.''}
	\Description{A visualization of the responses to the questions Q13 and Q14 is shown. For Q13 labeled ``found transparency helpful'', 0 strongly disagree and 0 disagree, 2 are neutral, 6 agree, 3 strongly agree. For Q14 labeled ``transparency perceived useful'', also 0 strongly disagree and 0 disagree, 1 is neutral, 7 agree, 3 strongly agree.}
	\label{fig:plot_transparency-helpful-useful}
\end{figure}

In our questionnaire, we therefore asked the eleven data owners (developers) among our participants to answer on a 5-point Likert scale if they found this transparency helpful and useful (Q13 \& Q14).
We find that data owners show very positive sentiment towards inverse transparency, judging it as beneficial (see \autoref{fig:plot_transparency-helpful-useful}).

To understand participants' individual perspectives, we followed up with a free text question about their experience with the provided transparency (Q15).
In general, participants expressed positive sentiment, explaining that the transparency ``makes me feel safe'' (O1) and that ``it is a great approach'' (O4) or ``a good tool'' (O10).
``I personally like to use it'' (O11), wrote one, with another expressing: ``I experienced it well'' (O5).
Notably though, one participant had a very different experience. They write: ``This transparancy [sic!] makes me even more aware of the monitoring, so im [sic!] not able to forget it.'' (O9). Thereby, they touch upon an important point. Many data usages may be benign, but being confronted with them may still have an influence.
For most of our participants, the additional transparency induced a feeling of safety, but some may find that it actually creates a sense of worry and a feeling of being monitored.
We reflect this thought in \autoref{sec:discussion}.

\subsection{RQ~\ref*{rq:influences-consumers}: Do participants consider inverse transparency to be capable to influence data consumers' data usage behavior?}

Receiving transparency over how data are used is judged as helpful and useful by data owners.
Yet, the question arises if this transparency influences data consumers, as it might elicit a feeling of ``being watched.''
Potential chilling effects could be desired (by preventing data misusage), or problematic (if legitimate usage of data is hindered).
Therefore, we asked participants about the potential influence of inverse transparency on data consumers' data usage behavior.

First, we asked the data owners ($n = 11$) if they consider inverse transparency a meaningful deterrent for potential data misusage (Q16).
We find just one participant disagreeing with the statement, with all others agreeing ($4/11$) or agreeing strongly ($6/11$).
In a free text answer, the critical respondent reveals that they consider the ``transparency [to be] high but not high enough to be bad'', judging the risk of misusage in our concrete setting to be low in general.
They add: ``I believe if usage of data is within good intention. It does not effect user behavior.''
This suggests that they might have misunderstood the question to be about our concrete scenario, not the theoretical possibility of misusage.

\begin{figure}[htbp]
	\centering
	\includegraphics[width=\linewidth]{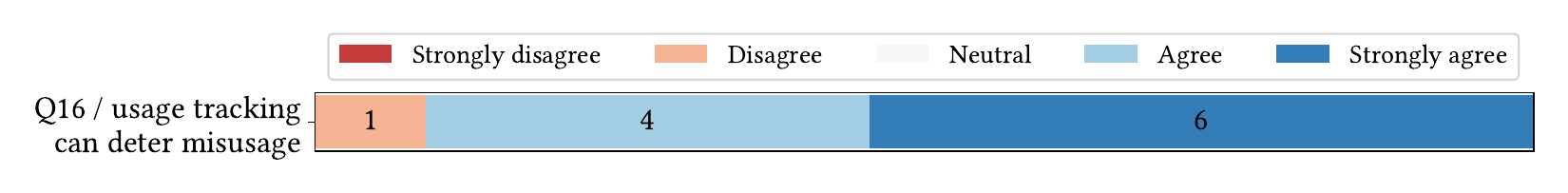}
	\caption{Results of the question to data owners if they think that inverse transparency can deter misusage. Data owners ($n = 11$) were asked to express their agreement on a five-point Likert scale (X-axis), from 1 for ``strongly disagree'' to 5 for ``strongly agree.''}
	\Description{A visualization of the responses to the question Q16 is shown, labeled ``usage tracking can deter misusage''. For it, 0 strongly disagree, 1 disagrees, 0 are neutral, 4 agree, 6 strongly agree.}
	\label{fig:plot_transparency-deters-consumers}
\end{figure}

Second, we asked the data consumers ($n = 4$) to elaborate if the monitoring of their data usage behavior influenced their actions (Q17).
Two consumers were sure, answering ``I believe so'' (C3) and ``Definitely'' (C2). Both explained that they think the monitoring reduced the number of their accesses.
``Without having to worry about potentially having to explain the frequency of my accesses to the data owners, I would likely have conducted the analyses more often'' (C3).
The third consumer responded that it only influenced them ``barely, but I can't be 100\% sure'' (C1), with the final participant being sure that the answer would be no: ``I would have acted the same'' (C4).
All of our participants could not know, but it is notable that opinions diverge so strongly. With two participants clearly worrying about the impression of their data usage on data owners, while another did not even consider it, other confounding factors may have been responsible.
For example, the team culture with regards to data-driven management may change the acceptance of the team lead utilizing analyses.
We conclude that inverse transparency is capable, but not guaranteed, to influence data consumers in their data usage behavior.

\subsection{RQ~\ref*{rq:valuable}: Is inverse transparency considered valuable in general by participants?}

Finally, we want to understand if participants would prefer inverse transparency over the status quo, which for them is the GDPR.
While data owners considered the provided transparency helpful and useful in our specific scenario (see \hyperref[rq:helpful]{RQ~\ref*{rq:helpful}}), they may still judge the value of inverse transparency differently if it was introduced to their workplace.
Therefore, we asked them for their agreement to four statements:

\begin{itemize}
	\item Inverse transparency improves upon the protection of the GDPR.~(Q18)
	\item I would prefer inverse transparency over just having the right to consent to or reject data usages outright.~(Q19)
	\item If my company offered me the choice, I would like to have access to data usage tracking.~(Q20)
	\item I would feel safer knowing how my data are accessed in detail.~(Q21)
\end{itemize}

In addition, we allowed participants to elaborate if they wanted to explain their responses (Q22).

\begin{figure}[htbp]
	\centering
	\includegraphics[width=\linewidth]{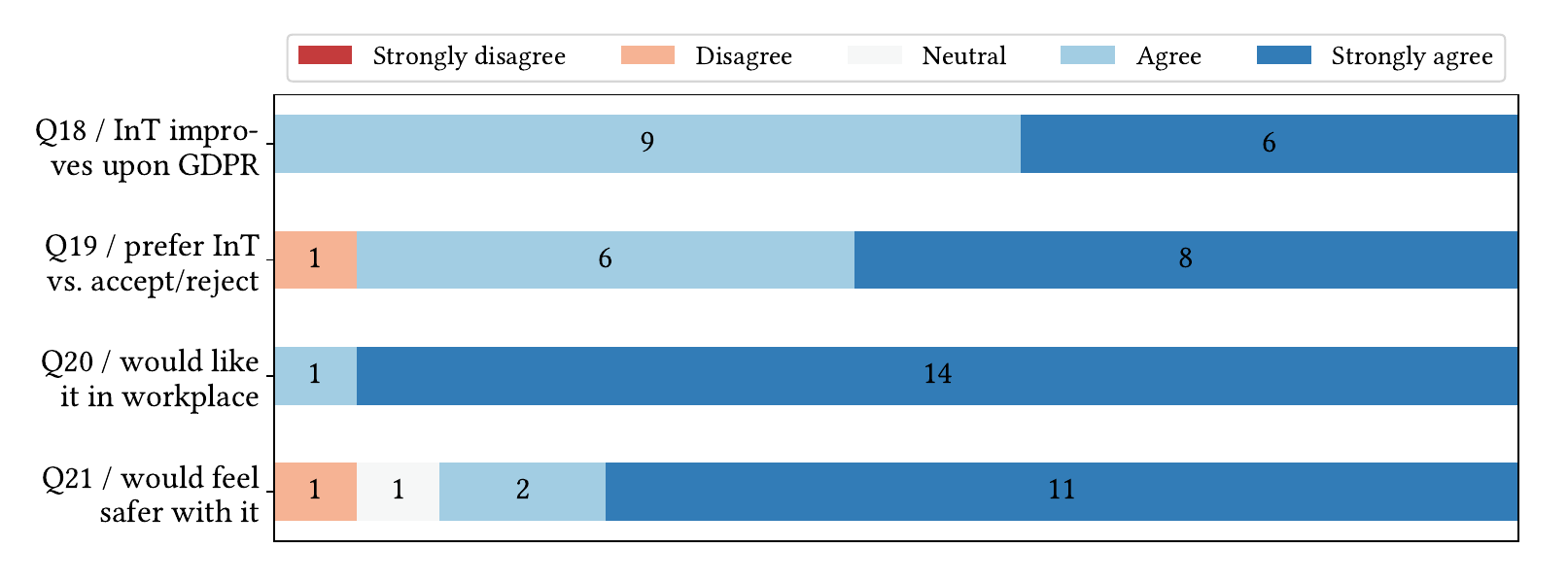}
	\caption{Results of the questions about inverse transparency generally (abbreviated as ``InT'' above). All participants ($n = 15$) were asked to express their agreement on a five-point Likert scale (X-axis), from 1 for ``strongly disagree'' to 5 for ``strongly agree.''}
	\Description{A visualization of the responses to the questions Q18 through Q21 is shown. For Q18 labeled ``InT improves upon GDPR'', 0 strongly disagree or agree or are neutral, 9 agree, and 6 strongly agree. For Q19 labeled ``prefer InT vs. accept/reject'', 0 strongly disagree, 1 disagrees, 0 are neutral, 6 agree, 8 strongly agree. For Q20 labeled ``would like it in workplace'', 0 strongly disagree and 0 disagree, 1 is neutral, 1 agrees and 14 strongly agree. For Q21 labeled ``would feel safer with it'', 0 strongly disagree, 1 disagrees, 1 is neutral, 2 agree and 11 strongly agree.}
	\label{fig:plot_inverse-transparency-valuable}
\end{figure}

Responses are overall very positive, with only one respondent disagreeing for questions Q19 and Q21, respectively.
For question Q21, the critical respondent elaborated that they ``would not trust that the access data provided to [them] is actually correct'' (O9). Possibly, if an integrity guarantee was given, their sense of safety could be improved.
Especially striking is the result for question Q20, with all participants agreeing and almost all ($14/15$) even expressing strong agreement. This suggests that inverse transparency is considered especially valuable by participants when imagining a workplace setting.
This sentiment is best summarized by the following quotes: ``I think that if implemented correctly, Inverse Transparency could be a game changer for privacy'' (C3), with another respondent adding: ``being able to track how my data is used is beneficial to myself, to transparancy [sic!] within the team and ultimately to the relationship between data owners and data consumers'' (O6).

\subsection{Conclusions}

Our study of the user perspective shows that inverse transparency is judged positively and as an improvement over the status quo.
All participants agreed that it would be a valuable addition to the workplace.
This suggests inverse transparency is capable to empower employees.
We furthermore find promising results regarding our goal of creating accountability and deterring misusage of data.
Data owners believed that the inverse transparency could influence data consumers' usage behavior, an impression that some data consumers confirmed for their case.
Finally, considering the purpose dimension of trust (see \autoref{sec:concept-trustworthiness}), participants judge inverse transparency as beneficial for themselves and their teams.
The introduction of inverse transparency is considered valuable for employees, which is reflected in that all participants would like it in the workplace.
This shows that the purpose of inverse transparency is perceived as benevolent and supportive of employees.

We conclude that inverse transparency is capable to meet our goals of empowering employees and creating accountability for data usages.
It is considered an improvement of the status quo, with the perceived benevolent purpose and felt benefits supporting trust in the concept.

\section{Discussion}
\label{sec:discussion}

At first glance, expecting a rethink of people analytics design may seem ambitious at best.
Yet, we have seen with the introduction of the GDPR how quickly software firms can adapt and update their tools.
We think the key driver for this speed were their customers---the companies buying these tools for their use.
To ensure their GDPR compliance, they expect their suppliers to build their software accordingly and integrate the necessary tools for, e.g., anonymization or deletion of user data.
Now, companies are faced with potent privacy legislation on the one hand, and privacy concerns of their employees~\cite[see][]{teebken2021privacy} on the other hand.
Ignoring these concerns can lead to anything from reactance~\cite{feng2019understanding} to dissatisfaction with the employer or psychological distress~\cite{bhave2020privacy}.
This means that it is in companies' own interest to address their employees' concerns.
Furthermore, where workers' councils are active, they may have a say in if analysis tools are bought, using their power to prevent invasive uses of employee data.
Inverse transparency could offer a solution for these scenarios. Not all data analyses are problematic, and some might even be beneficial for data owners. Tools built with inverse transparency by design can unlock this potential by empowering data owners with data sovereignty and holding data consumers accountable in case of data misusage.

The shift to develop with inverse transparency by design may have a more pronounced influence on software design than it first appears. We have deliberated the potential changes that can be envisioned when considering \emph{ambient usage} of data. Developers may need to change fundamental interaction paradigms as they include data owners as stakeholders in their design.
On the other hand, as we could see in our preliminary study of the developer perspective, inverse transparency may serve as an enabler for potentially privacy-invasive but useful features. Knowing that data owners will be able to supervise usage of their data may reduce the need to be overly cautious upfront, enabling the ethical development of innovative features.

Considering the users, providing inverse transparency is a valuable step but not sufficient on its own.
On the one hand, participants in our preliminary study of the user perspective showed a clear interest in inverse transparency.
They felt empowered and considered the provided transparency helpful and useful.
Yet, too much transparency~\cite{kizilcec2016much} or ``wrong'' ways to frame this transparency~\cite{rader2018explanations} may reduce trust in users.
One of our study participants confirmed this, noting that they felt more anxious due to the detailed insight into how their data were used.
In addition, if explanations or insights are too technical, this too can render them ineffective or even counterproductive~\cite{rawlins1994measuring, cramer2007user}.
Making the usage logs comprehensible for individuals is therefore an important part of providing transparency.
This can mean following users' mental model to create an intuitive tool or, if necessary, providing explanations and documentation~\cite{murmann2017tools}.
Moreover, we need to consider the possibility of habituation.
While our study spanned three months and therefore provides more than just a snapshot, it cannot predict the very long term effects of having access to relatively repetitive data usage information.
\citeauthor{karegar2020dilemma}~\cite{karegar2020dilemma} discuss the risk of habituation in the context of privacy notices, as it can reduce user attention.
To combat this, they suggest engaging users in different ways~\cite{karegar2020dilemma}.
For example, an automated anomaly detection system could be incorporated into the transparency dashboard to highlight unusual usage patterns.

At the same time, the effectiveness of inverse transparency systems in empowering employees depends on the sociotechnical context of the workplace.
The insights of people analytics can be important for management~\cite{davenport2010competing}, which may lead to external pressures for employees to contribute data~\cite{zieglmeier2022increasing} even if they perceive misusage.
\citeauthor{seberger2021empowering}~\cite{seberger2021empowering} find that, in such cases, technical mechanisms meant to empower users may not be sufficient.
Instead, if users perceive no alternative, they accept privacy violations more easily.
In fact, seemingly empowering mechanisms may then conversely even lead to resignation due to perceived personal responsibility.~\cite{seberger2021empowering}
This is a critical aspect especially for the workplace context, as the power asymmetry can reduce individual agency further.
The fear of being fired or facing other negative consequences precludes empowerment~\cite{chamorro2020can}.
It is therefore necessary to investigate how to actually empower employees to handle misusage of their data.
This should include investigating the trade-offs they face in their choices~\cite[Table~2]{sannon2022privacy}.
For example, technical solutions that ensure \emph{plausible deniability} may be essential to remove the pressure to conform~\cite{deng2011privacy}.

Finally, while our studies show promising results, they are preliminary and therefore potentially limited in their significance.
The use of university students as subjects, especially in a controlled environment, can threaten the validity of studies researching the workplace.
Students have limited experience with working environments, lacking knowledge of its broader context and influencing factors.
Additionally, the inherent complexity of the workplace as well as various confounding factors cannot be fully mirrored in an artificial setting.
For our purposes, though, this seeming limitation was instead a sought-after property.
We worked closely with our industry partner on a conceptual level to develop realistic use cases for inverse transparency.
For our experimental studies, meanwhile, we specifically chose to model a fully controlled workplace-like environment with computer science students instead.
Experiments with students can be preferable when testing initial hypotheses~\cite[p.~739]{sjoberg2005survey}.
Computer science students specifically are judged as sensible stand-ins for professionals~\cite{tichy2000hints}.
Our proposal to introduce inverse transparency in the workplace means a fundamental change in the work and interactions of employees.
Conducting our studies in an existing workplace could have triggered confounding change management issues, most notably individual's resistance to change~\cite{bovey2001resistance}.
This is an important factor limiting the success of change initiatives~\cite{kuhlman2021will}.
Furthermore, external influences such as time pressure from other projects could have influenced our results~\cite{nan2009impact}.
Removing these factors was necessary to support internal validity~\cite{fiske2005laboratory} and establish causality~\cite{katok2018designing}.
This allows us to cleanly link the study results to our intervention of inverse transparency by design.
Additionally, it is infeasible to fully control a real world working environment, including the work tasks and utilized tools, continuously for multiple months.
This level of control was essential, though, especially for our study of the user perspective.
By completely aligning work tasks and incentives with our study goals, we improve construct validity compared to a less controllable real world environment~\cite{frechette2015laboratory}.
At the same time, we recognize our studies as preliminary, as their artificial nature could reduce external validity.
A promising next step may therefore be to explore if confounding factors in a real workplace influence individuals' perceptions.
Given those limitations, our results show that inverse transparency by design can be practical from a technical standpoint and is experienced as beneficial by our study participants.
We consider these insights promising and a sign that inverse transparency by design has the potential to be an important factor in accepted and responsible people analytics.


\begin{acks}
This work was supported by the \grantsponsor{BMBF}{German Federal Ministry of Education and Research (BMBF)}{https://www.bmbf.de/} under grant no. \grantnum{BMBF}{5091121}.
We thank Antonia Maria Lehene for her help in creating and validating our user trust questionnaire.
Furthermore, we thank the anonymous reviewers for their helpful comments, which greatly improved the final paper.
\end{acks}


\bibliographystyle{ACM-Reference-Format}
\bibliography{references}

\clearpage
\appendix

\section{Addendum for study B}

In the following, find a screenshot of the transparency dashboard (\ref{sec:appendix-screenshot-clotilde}) and a list of the questionnaire questions (\ref{sec:appendix-questionnaires}).

\subsection{The transparency dashboard}
\label{sec:appendix-screenshot-clotilde}

\begin{figure}[h!]
	\centering
	\includegraphics[height=0.79\textheight]{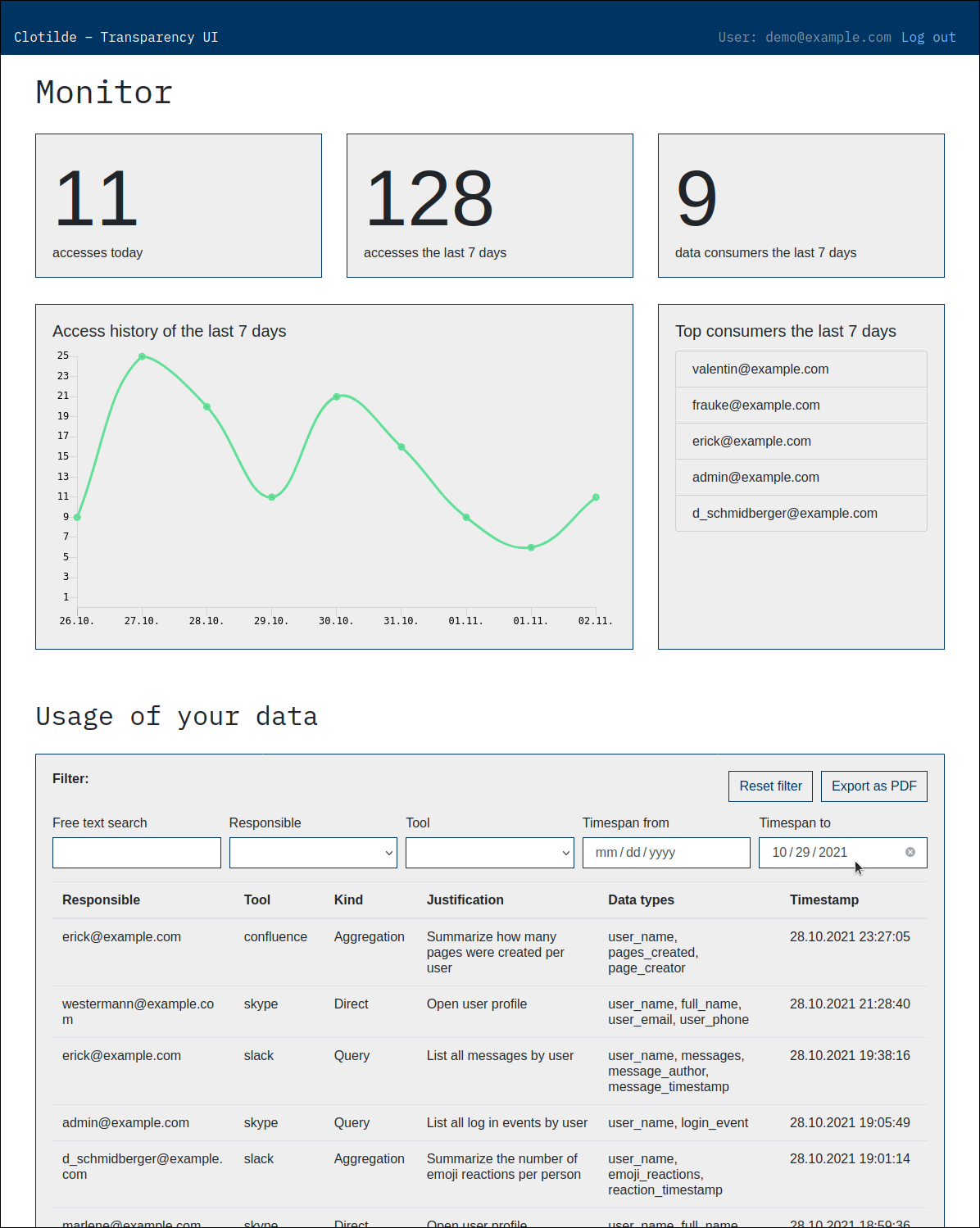}
	\caption{Screenshot of the transparency dashboard used in study B. The top half summarizes the data usages recorded in the last seven days. The bottom presents a detailed list of the logged data usages for data owners to inspect. The email addresses and usage data shown are artificial and for illustration purposes only.}
	\Description{The transparency dashboard is shown in full. Going over it top to bottom, left to right: It starts with a header bar in blue, containing a title ``Clotilde – Transparency UI'' and on the right ``User: demo@example.com'' with a button ``Log out''. Then the title ``Monitor''. Then five boxes with varying info. Three in the first row with ``11 accesses today'', ``128 accesses the last 7 days'' and ``9 data consumers the last 7 days''. Two in the second row with the first showing a graph titled ``Access history of the last 7 days'', the second a list of ``top consumers the last 7 days''. Below it a title for the second section, namely ``Usage of your data''. Following a last box with a set of filter elements followed by a table. Filter elements include a free text search and timespan selection, and the table consists individual log items with ``Responsible'', ``Tool'', ``Kind'', ``Justification'', ``Data types'' and ``Timestamp''. For example, one line reads: ``erick@example.com'' with tool ``confluence'', kind ``aggregation'', justification ``Summarize how many pages were created per user'', data types ``user\_name, pages\_created, page\_creator'', and timestamp ``28.10.2021 23:27:05''.}
	\label{fig:screenshot_clotilde-overview}
\end{figure}

\subsection{Questionnaire questions}
\label{sec:appendix-questionnaires}

In the following, find the questions answered by participants in study B.

\subsubsection{RQ 1 -- Unified dashboard}

\begin{itemize}
	\item Q1--Q10 (Likert scale): System usability scale (questions in~\cite{brooke1986system})
	\item Follow-up free text questions
	\begin{itemize}
		\item Q11: Did you enjoy using Clotilde\footnote{The name of the transparency dashboard.}? Why?
		\item Q12: What could be changed to improve your experience? What did you miss?
	\end{itemize}
\end{itemize}

\subsubsection{RQ 2 -- Inverse transparency experienced as beneficial}

\begin{itemize}
	\item Likert questions
	\begin{itemize}
		\item Q13: I found the additional transparency helpful.
		\item Q14: Having more transparency over data usages was useful.
	\end{itemize}
	\item Q15 (free text): How did you experience the provided transparency? What would have changed your experience?
\end{itemize}

All questions of this RQ were for data owners.

\subsubsection{RQ 3 -- Inverse transparency can influence data consumers}

\begin{itemize}
	\item Data owners -- Q16 (Likert scale): I think that the usage tracking could deter data users from misusing my data.
	\item Data consumers -- Q17 (free text): Would you have acted differently in case your accesses weren't monitored?
\end{itemize}

\subsubsection{RQ 4 -- Inverse transparency considered valuable}

\begin{itemize}
	\item Likert questions
	\begin{itemize}
		\item Q18: Inverse Transparency improves upon the protection of the GDPR (DS-GVO).
		\item Q19: I would prefer Inverse Transparency over just having the right to consent to or reject data usages outright.
		\item Q20: If my company offered me the choice, I would like to have access to data usage tracking.
		\item Q21: I would feel safer knowing how my data are accessed in detail.
	\end{itemize}
	\item Q22 (free text): Optional: Any further comments?
\end{itemize}

\received{July 2022}
\received[revised]{January 2023}
\received[accepted]{March 2023}

\end{document}